\documentstyle[12pt]{article}
\newcommand{\T}{\mbox{\rm Tr}}
\newcommand{\MPL}{Mod. Phys. Lett. {\bf A}}
\newcommand{\NPB}{Nucl. Phys. {\bf B}}

\newcommand{\PLB}{Phys. Lett. {\bf B}}
\newcommand{\IJM}{Int. Jour. Mod. Phys. {\bf A}}

\begin{document}

%%%%%%%%%%%%%%%%%%%%%% TITLE PAGE %%%%%%%%%%%%%%%%%%%%%%%%%%%%%%%%%%%%%%%
\begin{flushright}
hep-th/9706223 \\
BROWN-HET-1051  \\
June 1997
\end{flushright}     
\begin{center}
{\Large 
{\bf Large ${\mathbf N}$ WZW Field Theory Of ${\mathbf N=2}$ Strings
\footnote{
supported by 
DOE under grant DE-FG0291ER40688-Task A.\\
email address: antal,nunes@het.brown.edu, mm@barus.physics.brown.edu}}}

Antal Jevicki, Mihail Mihailescu and Jo\~ao P. Nunes\\

Department of Physics\\
Brown University\\
Providence RI 02912

{\bf Abstract}
\end{center}

We explore the quantum properties of self-dual gravity formulated as a
large $N$ two-dimensional WZW sigma model. 
Using a non-trivial classical background, we show that a ${\rm (2,2)}$
space-time is generated. The theory contains an infinite series of higher 
point vertices.
At tree level we show that, in spite of the presence of higher 
than cubic vertices, the on-shell 4 and higher point functions vanish, 
indicating 
that this model is related with the field theory of closed $N=2$ strings. 
We examine the one-loop on-shell 3-point amplitude and show that it is 
ultra-violet finite.

\newpage
%%%%%%% SECTION 1 INTRODUCTION%%%%%%%%%%%%%%%%%%%%%%%%%%%%%%%%%%%%%%%%%%%%
\section{Introduction}

String theory through its worldsheet description exhibits
notable finiteness properties. Its target space description in terms of
field theory is then of major interest both with regard to 
nonperturbative studies, but more recently also as a building block of higher 
dimensional extended objects. One of the simplest models is given by the 
$N=2$ string \cite{dadda}--\cite{L}. 
This theory, due to its solvability, has been 
extensively studied since its discovery \cite{dadda,dadda1} and served as a 
laboratory for acquiring deeper insight into the structure of string theory. 

The main features of the $N=2$ string
are the appearance of a four-dimensional ${\rm (2,2)}$ target space and the 
vanishing 
of all higher point amplitudes \cite{OV}--\cite{L}.
Its field theoretic  description seems to be given
by versions of self-dual Yang-Mills theory and self-dual gravity.
This field theoretic representations are subjects of major interest and
are at present known to contain a  number of puzzles. In particular, at 
the quantum
level there is as yet no clear way of defining self-dual gravity, the theory 
is accompanied by infinities and it is in fact string theory that may
suggest the proper definition of these four-dimensional theories.
The interest in these theories has increased recently \cite{J,KM}
since through the worldsheet from target
space mechanism \cite{G,KM} they are promising candidates for building 
blocks for a worldsheet description of membranes.
Indeed, a field theoretic formulation \cite{J} of
the membrane matrix model was seen to be given by an $SU(\infty)$ WZW
theory in two dimensions. Since the additional worldsheet coordinates appear 
 from the large $N$ color degrees of freedom \cite{Ho,FIT} one finds a 
similarity
 with various sigma model descriptions of self-dual gravity 
\cite{Pl}--\cite{PP}. 
This approach to self-dual gravity has so far been explored
strictly at the \underline{classical level}: the equations of motion when 
written in 
terms of currents lead to Pleba\'nski's heavenly equations \cite{Pl}, with 
different sigma models being classicaly equivalent. One might envision a
direct quantization of the cubic Pleba\'nski 
action, but some studies of non-abelian duality transformations for 
sigma models \cite{FJ}
would indicate that this could exhibit problems, such as the wrong beta 
function.

In this paper, we approach the problem of quantizing self-dual gravity and 
developing a field theory of $N=2$ closed strings based directly
on a (perturbed) WZW model represented as a non-linear field theory with an 
infinite series of higher point vertices.
The fact that the WZW model
possesses remarkable quantum properties (such as vanishing  
beta function) offers the hope that the corresponding gravitational 
theory will be equally well defined. We expect that the infinite sequence of 
higher point vertices is likely to be relevant for its finiteness. 
The first question at hand 
is then to verify the vanishing of higher than 3-point on-shell amplitudes 
which is what we do in this paper. We also proceed to study the theory at 
one-loop and 
demonstrate the vanishing of ultra-violet divergences in the computation of 
the on-shell 3-point amplitude. We establish a correspondence of our model
with a particular large $N$ limit of four-dimensional WZW theory 
\cite{NaS,LMNS}.

%%%%%%%%%%%%%%%%%%%%% SECTION 2 %%%%%%%%%%%%%%%%%%%%%%%%%%%%%%%%%%%%%%%%%%%%

\section{The Model}

We will consider a two dimensional sigma model defined in a space-time with 
Lorentzian signature ${\rm (1,1)}$, light-cone coordinates 
$x^{+}=x^{1}+x^{2}$, $x^{-}=x^{1}-x^{2}$ and metric 
$ds^{2}_{2d}=dx^{+}dx^{-}$. 
The model will be a chiral model with a Wess-Zumino term,
where the coefficient of the chiral term will be perturbed from its
conformal fixed point value, so that our model can be seen as a perturbed 
WZW model at level $K$,
\begin{equation}
S(g)=\frac{(1+\epsilon)K}{4\pi}\int dx^{+}dx^{-}\T
(\partial_{+}\,g\partial_{-}g^{-1}) + \frac{K}{12\pi} \Gamma_{WZW}(g),
\label{2.2}
\end{equation}
where $g$ is the group valued field,
$\epsilon$ is a real parameter and $\Gamma(g)$ is the Wess-Zumino term.
When $\epsilon =0$ we obtain the usual WZW model fixed point. The group $G$
remains unspecified for now, but shortly we will take the Lie algebra to be 
some large $N$ limit of $su(N)$.

The classical equations of motion will be given by
\begin{equation}
\frac{\epsilon}{1+\epsilon}\;\partial_{+}(g^{-1}\partial_{-}g)+
\frac{2+\epsilon}{1+\epsilon}\;\partial_{-}(g^{-1}\partial_{+}g)=0.
\label{2.3}
\end{equation}

As long as $\epsilon\neq -2,-1,0$, one can define $Q$ by
$$
\frac{2+\epsilon}{1+\epsilon}\;g^{-1}\partial_{+}g=Q^{-1}\partial_{+}Q
\;\;\;\; {\rm and} \;\;\;\;
\frac{\epsilon}{1+\epsilon}\;g^{-1}\partial_{-}g=Q^{-1}\partial_{-}Q
$$
where $Q$ is a classical solution of the pure chiral model, see for example
\cite{Z},
$$
\partial_{+}(Q^{-1}\partial_{-}Q)+\partial_{-}(Q^{-1}\partial_{+}Q)=0.
$$
To make the connection between the two-dimensional sigma model (\ref{2.2})
and four-dimensional self-dual gravity, we start by
expanding the field $g$ aroung a specific classical configuration. For reasons
that will become clear shortly, this classical background will be
\begin{equation}
Q^{-1}\partial_{+}Q=\frac{{\hat q}}{\Lambda}+
\frac{x^{-}}{2\Lambda^{2}};\;\;\;\;\;\; 
Q^{-1}\partial_{-}Q=\frac{{\hat p}}{\Lambda}-
\frac{x^{+}}{2\Lambda^{2}},
\label{2.4}
\end{equation}
where the matrices ${\hat q}$ and ${\hat p}$ satisfy 
$[{\hat q},{\hat p}] =1$ and $\Lambda$ is an arbitrary parameter.
In the context of our approach, what is important about ${\hat q}$ and 
${\hat p}$ is that in the large $N$ limit they become canonical variables
$q$ and $p$, possibly after some rescaling by factors of $N$.
The corresponding $g$ will be given by
\begin{equation}
g_{0}^{-1}\partial_{+}g_{0}=\frac{1+\epsilon}{2+\epsilon}\,
(\frac{{\hat q}}{\Lambda}+\frac{x^{-}}{2\Lambda^{2}});
\;\;\;\;\;\; 
g_{0}^{-1}\partial_{-}g_{0}=\frac{1+\epsilon}{\epsilon}\,
(\frac{{\hat p}}{\Lambda}-\frac{x^{+}}{2\Lambda^{2}}).
\label{2.5}
\end{equation}

We will now expand around this classical background. We take 
$N\rightarrow\infty$
such that the fields will take values in the infinite dimensional Poisson 
algebra of functions on some two-dimensional space $\Sigma$, 
$sdiff(\Sigma )$. The Lie bracket is then given by the Poisson bracket
\begin{equation}
\{ f,g\} = \partial_{q}f\partial_{p}g-\partial_{p}f\partial_{q}g
\label{2.1}
\end{equation}
where $f$,$g$ are functions of the coordinates $q$,$p$ on $\Sigma$.
The trace becomes the integration over the $su(\infty)$ ``color''
variables $q$ and $p$.
It is well known that this Lie algebra is closely related to the large $N$
limit(s) of $SU(N)$ \cite{Ho}. The quantum fluctuations will be 
described by the field $\omega$, which will take values in the Poisson 
algebra, where
\begin{equation}
g=g_{0}\exp (\omega).
\label{2.6}
\end{equation}

From the Polyakov-Wigman 
formula \cite{PW,LS} and after rescaling the field $\omega$ to normalize the 
kinetic term we obtain\footnote{The $x^{+}$ and $x^{-}$ dependent terms in 
the classical background currents (\ref{2.5}) give vanishing contributions 
to $S(\omega)$.}
%\newpage
\begin{eqnarray}
\nonumber
S(\omega )=\frac{1}{2}\int dx^{+}dx^{-}\T (
\partial_{p}\omega\partial_{-}\omega -\partial_{q}
\omega\partial_{+}\omega) 
\phantom{777777777777777777777777777}
\\
\nonumber
+\frac{1}{3!} \kappa\int dx^{+}dx^{-} 
\T (\omega\{\partial_{p}\omega,\partial_{-}\omega\}
-\omega\{\partial_{q}\omega , \partial_{+}\omega\} ) 
\phantom{777777777777777}
\\
\nonumber
+\frac{1}{4!}\kappa^{2}\int dx^{+}dx^{-} \T 
(\partial_{q}\omega\{\{\partial_{+}\omega ,\omega\} ,\omega\}
-\partial_{p}\omega\{\{\partial_{-}\omega,\omega\} ,\omega\})+\cdots
\phantom{7777777}
\\
+\Lambda\int dx^{+}dx^{-}\T (-\partial_{+}\omega\partial_{-}\omega
-\frac{2}{3!}\kappa\omega\{\partial_{+}\omega ,\partial_{-}\omega\}
+ \frac{2}{4!}\kappa^{2}
\omega\{\{\partial_{-}\omega ,\omega\},\partial_{+}\omega\}) + \cdots
\label{2.7}
\end{eqnarray}
where the coupling constant 
$\kappa = \sqrt{\frac{4\pi\Lambda }{K(1+\epsilon )}}$.
We see that we generate cubic, quartic and higher point vertices for the 
field $\omega$. Moreover, the two-dimensional propagator 
$\partial_{+}\omega\,\partial_{-}\omega$, together with other terms without
derivatives in the color directions, is multiplied by factor of $\Lambda$.
In the limit $\Lambda\rightarrow 0$ these terms disappear and the remaining
quadratic terms in $\omega$ define 
a four-dimensional-looking propagator for a metric of ${\rm (2,2)}$ signature
\footnote{Notice that the fact that we obtain signature ${\rm (2,2)}$
is closely related with the choice of sign in the classical background
(\ref{2.4}).}
\begin{equation}
ds^{2}_{4d}=dx^{+}dq-dx^{-}dp.
\label{2.8}
\end{equation}

We will take $\Lambda\rightarrow 0$ for the remainder of the paper, 
defining a double-scaling limit where the level $K$ and the parameter 
$\epsilon$ are such that $\kappa$ is finite and nonzero. 
The cubic vertex of order $\kappa$ appearing in (\ref{2.7}) represents an 
$SO(2,2)$ Lorentz transformation of the cubic 
vertex $\omega\{\partial_{+}\omega ,\partial_{-}\omega\}$, whose
equation of motion is one of the forms of the Pleba\'nski equation
\footnote{We note that the Pleba\'nski equation is not $SO(2,2)$ invariant, 
at least manifestly. For a discussion of the Lorentz symmetries of self-dual
gravity see \cite{PBR}.} for four-dimensional self-dual gravity. 
Its amplitudes \cite{PA} are closely related to  the ones of the closed $N=2$ 
string \cite{OV}--\cite{L}.
However, we note that in the present model 
there is in addition an infinite series of higher point vertices.

When we compare with the $N=2$ string field theory of \cite{OV} we will
identify $\kappa$ with the closed string coupling constant. We note that at 
this point we can analytically continue in $x^{+},x^{-},q,p$ so that we get 
the flat K\"ahler metric 
\begin{equation}
ds^{2}_{4d}=dyd{\bar y}-dzd{\bar z}
\label{2.10}
\end{equation}
where $y,{\bar y},z,{\bar z}$ are complex coordinates corresponding to 
$x^{+},q,x^{-},p$ respectively. The field $\omega$ is then related with 
deformations of the flat K\"ahler potential \cite{OV}.
We will use the notation in (\ref{2.10}) in the remainder of the paper.

Since
in our approach we also generate an infinite series of higher point vertices,
in the next section, we will study the amplitudes coming from these vertices
and will show that they are also compatible with the $N=2$ closed string. 
However,
the fact that higher point vertices are present, and the close relation
of this theory with the WZW model, leads us to believe that it will 
have nice properties beyond the one-loop level. Moreover, the action 
(\ref{2.2}) provides a systematic expansion for determining these higher 
point vertices.
It is hoped that the fact that two of the four dimensions appear as
large $N$ color variables in disguise, may shed some light into the problem
of matching string and field theory amplitudes at one-loop level
\cite{OV,BGI,M,L}. In fact, the large $N$ approach suggests a possible 
infra-red regulator for the momenta along ${\bar y},{\bar z}$. There is also 
the possibility that some specific choice of $\Sigma$, for example a torus 
$T^{2}$, will regulate these momenta even at infinite $N$.
 
%%%%%%%%%%% SECTION 3 %%%%%%%%%%%%%%%%%

\section{The Amplitudes}

In the next two sub-sections we will examine the field theory amplitudes of
(\ref{2.2}), always with $\Lambda\rightarrow 0$. We will first look at the 
tree level amplitudes, where we will show that the 4 and 5-point 
functions vanish on-shell. We will also comment on the relation between 
(\ref{2.2}) and four-dimensional self-dual Yang-Mills theory.
We will then look at the one-loop on-shell 3-point function and show that 
the new 4 and 5-point vertices do not contribute to it.

\subsection{Tree Level Amplitudes}

The metric is given by (\ref{2.10}) and the corresponding components of the 
momenta $k$ will be $k_{y},k_{{\bar y}},k_{z},k_{{\bar z}}$. The inner 
product will then be given by 
$k_{i}\cdot k_{j}=\overline{k_{j}\cdot k_{i}}=k_{iy}k_{j{\bar y}}-
k_{iz}k_{j{\bar z}}$.

Following \cite{OV,PA} we introduce the kinematical quantities
\footnote{The quantity ${\tilde c_{ij}}^{2}$ of \cite{OV} is given by
$a_{ij}{\bar a}_{ij}$. When $k_{i}$,$k_{j}$ and $(k_{i}+k_{j})$ are on-shell 
this becomes $c_{ij}^{2}$.},
\begin{eqnarray}
\nonumber
a_{ij}=-a_{ji}=k_{iy}k_{jz}-k_{jy}k_{iz}\;; \phantom{77777}
{\bar a}_{ij}=-{\bar a}_{ji}=
k_{i{\bar y}}k_{j{\bar z}}-k_{j{\bar y}}k_{i{\bar z}}\;;\\
k_{ij}=k_{i}\cdot k_{j}\;; \phantom{7777} s_{ij}=s_{ji}=k_{ij}+k_{ji}\;;
\phantom{7777} c_{ij}=-c_{ji}=k_{ij}-k_{ji}.
\label{3.1}
\end{eqnarray}

The propagator then becomes
\begin{equation}
\Delta (k,-k) = \frac{1}{k\cdot k} = \frac{1}{2s_{kk}}.
\label{3.45}
\end{equation}

The tree level 3-point function receives contributions only from the cubic 
vertex of (\ref{2.2}) and, on-shell, it is simply 
$V_{3}=\kappa\, c_{13}{\bar a}_{13}$. In \cite{PA}, Parkes shows that this 
can be obtained from the usual
on-shell Pleba\'nski vertex, which is $a_{13}{\bar a}_{13}$,
by an $SO(2,2)$ transformation. In our approach, since the ${\bar y}$ and 
${\bar z}$ coordinates are ``color'' variables, Lorentz transformations
are related to Lie algebra redefinitions of the field $\omega$, for example 
through commutators with $q$ and $p$. For example, if we were using exactly
the WZW model, the classical background would be a product of one 
anti-holomorphic term on the left and one holomorphic term on the right.
We could choose to define the quantum fluctuations either on the left or right,
or even in between these two terms. The resulting four-dimensional field
theory would have looked 
different, and the different quantum fields would be related by Lie algebra 
operations which would be connected with Lorentz transformations. Of course,
these quantum theories would be essentially equivalent. 

The $N=2$ string 
tree level 3-point amplitude is given by $g_{str}\,c_{13}^{2}$. When we
compare it with our field theory amplitude, while keeping in mind
the issue of Lorentz transformations, we conclude that $\kappa = g_{str}$.
In \cite{PA}, it is also verified that the tree level on shell 4 and 
5-point functions coming from this cubic vertex vanish. Therefore, when 
checking
these amplitudes for our model (\ref{2.2}), we need not consider graphs 
where the cubic term enters alone. 

We will now show directly that the tree level on shell 4-point function 
vanishes.
Later, we will relate our model with self-dual Yang-Mills and show why 
the 5 and higher on-shell tree level amplitudes also vanish.
The only graph we need to examine is the one with a single 4-point vertex,
since the contribution of the cubic term gives zero. The term where the legs 
1 and 2 are contracted, denoted by (12,34), will be given by
\begin{equation}
{\bar a}_{12}{\bar a}_{34}[k_{14}-k_{13}+k_{23}-k_{24}]
\label{3.2}
\end{equation}
If we put this together with (34,12) we get
\begin{equation}
{\bar a}_{12}{\bar a}_{34}[s_{14}-s_{13}+s_{23}-s_{24}]=
2{\bar a}_{12}{\bar a}_{34}(s_{23}-s_{13}),
\label{3.3}
\end{equation}
where we have used momentum conservation and the on-shell property $s_{11}=0$,
etc. The final result is obtained by summing the remaining four terms 
(13,24),(24,13),(14,23) and (23,14), which after using momentum conservation 
and the on-shell condition, gives
\begin{equation}
V_{4}\sim\kappa^{2}[{\bar a}_{12}{\bar a}_{13}s_{23}+{\bar a}_{13}{\bar a}_{23}s_{12}
-{\bar a}_{12}{\bar a}_{23}s_{13}].
\label{3.4}
\end{equation}
But, as a consequence of the highly constrained kinematics in ${\rm (2,2)}$
signature, the term in brackets in (\ref{3.4}) vanishes when all momenta
are on-shell \cite{OV,PA}. Therefore, the total on-shell tree level 4-point
amplitude for (\ref{2.2}) vanishes.

One could proceed and 
check that the 5 and higher point functions vanish on-shell. 
But let us show that the double scaling limit 
($\Lambda\rightarrow 0$, $K(1+\epsilon)\rightarrow 0$) of the two-dimensional 
WZW model that we are 
considering, is related in a specific way to a large $N$ limit of
four-dimensional self-dual Yang-Mills theory. The vanishing of the higher 
point amplitudes will then be seen as a consequence of this relation.

Consider the self-dual 
Yang-Mills equations
\begin{equation}
F_{{\bar \mu}{\bar \nu}}=F_{\mu\nu}=0\;,\phantom{777}
\eta^{\mu {\bar \nu }}F_{\mu {\bar \nu }}=0,
\label{3.5}
\end{equation}
in the gauge where $A_{\bar \mu }=0$. The first equation in (\ref{3.5}) is 
solved by $A_{\mu }=g^{-1}\partial_{\mu }g$ so that the equation of motion,
Yang's equation, reads
\begin{equation}
\eta^{\mu {\bar \nu }}\partial_{\bar \nu }(g^{-1}\partial_{\mu }g)=0.
\label{3.6}
\end{equation}
We now consider the Donaldson-Nair-Schiff action \cite{NaS,LMNS}
\begin{equation}
S=\frac{i}{4\pi}\int d^{4}x\T (g^{-1}\partial^{\mu}gg^{-1}\partial_{\bar \mu}g)
+\frac{i}{12\pi}\int_{M_5} \omega\wedge\T(g^{-1}dg)^{3},
\label{3.7}
\end{equation}
where $\T$ denotes the trace on the Lie algebra of the gauge group, $G(N)$.
Parametrizing $g(x^{\mu},x^{\bar \mu})=\exp ({\hat \phi})$, one expands the 
action in powers of the field ${\hat \phi}$. In the limit $N\rightarrow\infty$,
we have a six-dimensional non-linear scalar field theory where the matrix field
${\hat \phi}(x^{\mu},x^{\bar \mu})$ becomes a scalar field 
$\phi(x^{\mu},x^{\bar \mu},q,p)$, where $q,p$ are the large $N$ color 
variables \cite{FIT}.
If we reduce by identifying $x^{\bar 1}$ with $q$ and $x^{\bar 2}$ with $p$,
that is if we impose
\begin{equation}
(\partial_{\bar 1}-\partial_{q})\phi = 
(\partial_{\bar 2}-\partial_{p})\phi =0,
\label{3.100}
\end{equation}
we obtain a four-dimensional theory. We mention that in the Leznov-Parkes gauge
this was seen to lead to the second heavenly equation of self-dual gravity 
\cite{PP}. In our case, we have a non-linear theory since the vertices follow 
from the four-dimensional WZW action evaluated in the large $N$ limit.
Expanding (\ref{3.7}), we have after some algebra
\begin{eqnarray}
\nonumber
S=\int d^{4}x\{\frac{1}{2}\phi\partial^{\mu}\partial_{\bar \mu}\phi
+\frac{1}{3!}\eta^{\mu{\bar \mu}}\varepsilon^{{\bar \rho}{\bar \nu}}
\phi\partial_{\mu}\partial_{{\bar \rho}}\phi\partial_{{\bar \mu}}
\partial_{{\bar \nu}}\phi
+\frac{1}{4!}\eta^{\mu {\bar \mu}}\varepsilon^{{\bar \lambda}{\bar \nu}}
\varepsilon^{{\bar \rho}{\bar \sigma}}\partial_{\mu}\partial_{{\bar \lambda}}
\phi
\partial_{{\bar \nu}}\phi\partial_{{\bar \mu}}\partial_{{\bar \rho}}\phi
\partial_{\sigma}\phi \\
+\frac{1}{5!}\varepsilon^{{\bar \rho}{\bar \sigma}}
\varepsilon^{{\bar \lambda}{\bar \nu}}\varepsilon^{{\bar \zeta}{\bar \xi}}
\partial_{{\bar \mu}}
\partial_{{\bar \rho}}\phi\partial_{{\bar \sigma}}\phi
\partial_{{\bar \zeta}}(\partial_{\mu}\partial_{{\bar \lambda}}
\phi\partial_{{\bar \nu}}\phi )\partial_{{\bar \xi}}\phi+\cdots\}
\phantom{7777}
\label{3.8}
\end{eqnarray}

These terms agree, to this order, with the vertices of the double scaling 
limit of our two-dimensional model (\ref{2.2}). This correspondence can 
now be used to conclude about the tree level on-shell amplitudes of our 
model from those of self-dual Yang-Mills theory. The later represents 
(at finite $N$) a field theory of open $N=2$ strings with Chan-Paton factors
for the gauge group, say $U(N)$ \cite{M}. These amplitudes are given in a 
factorized sum over non-cyclic permutations 
$$
\sum_{\sigma}\T (T_{\sigma_{1}}T_{\sigma_{2}}\cdots T_{\sigma_{n}})
S_{n}(k_{1},...,k_{n}).
$$
In momentum space, the reduction (\ref{3.100}) is performed by identifying
the conjugate momenta $k_{q}=k_{{\bar 1}}$ and $k_{p}=k_{{\bar 2}}$. 
At tree level, by momentum conservation at the vertices, if these relations
are imposed on the external momenta they will be preserved throughout the 
Feynman graphs.
Then, the vanishing of the open $N=2$ string amplitudes for $n\geq 4$, 
implies
the vanishing of the corresponding $S_{n}$'s and also of the amplitudes for 
the model (\ref{2.2}),
indicating that it is indeed an appropriate field theory of the closed $N=2$ 
string, at least at tree level.
At loop level however, there are integrations of the momenta along the loops,
and this argument does not apply. We therefore proceed with a direct 
calculation.

\subsection{The One-Loop 3-Point Function}

In the usual cubic action for the field theory of the closed $N=2$ string, 
the one-loop 3-point amplitude \cite{BGI} is less infra-red divergent than the
corresponding string amplitude, while it is also ultra-violet finite. 
In our case, with the action in (\ref{2.2}), we will obtain similar results.
However, as we mentioned before, the underlying $2+2$ structure, with two
dimensions coming from color, may if further explored solve this problem.

At one-loop, the 3-point amplitude may receive contributions from several 
types of graphs. This is to be contrasted with the cubic theory where only one
type of graph enters. Although, as we will see below, only this graph 
contributes also in our case, it is tempting to conjecture that the higher
point vertices in (\ref{2.2}) will be important to ensure good properties of 
the theory at more than one loop level. Indeed, that is the case for the usual
two-dimensional WZW model. Let us examine these different diagrams each at a
time. 

The diagram with one 3-point tadpole vanishes because the cubic vertex 
is zero when two of the momenta at the vertex are equal. To examine the 
remaining graphs it is convenient to introduce Schwinger parameters.
We will use them to find the symmetry properties of the various terms,
and to find wish terms will vanish. However, we note that the fact that 
we are in signature ${\rm (2,2)}$ makes the definition of these
integrals a subtle one. 

The remaining Feynman diagrams contain potentially ultra-violet divergent 
terms with a naive behaviour $\sim M^{6}$,.., $\sim M^{2}$,
for an UV cut-off $M$, but as we will see these terms are in fact zero.
We have the diagram with one 5-point vertex and one propagator. After 
considering all possible leg permutations entering the vertex, we obtain,
with $\epsilon$ a regulator for the $\alpha$ integral,
\begin{equation}
\sum_{\sigma\in S_{3}} \int_{0}^{\infty} d\alpha \int d^{4}p
\exp(-\alpha\epsilon)\exp(i\alpha p^{2})[-2{\bar a}_{\sigma{1}p}+
{\bar a}_{\sigma_{3}p}]{\bar a}_{\sigma_{1}p}{\bar a}_{\sigma_{2}\sigma_{3}}
c_{\sigma_{2}\sigma_{3}} = 0.
\label{3.9}
\end{equation}
This vanishes
because ${\bar a}_{ip}$ contains only the $p_{\bar y}$ and $p_{\bar z}$
components of $p$ such that when we perform the gaussian integral over 
$d^{4}p$ we obtain zero. Similar arguments show that the graphs with one 
3-point and one 4-point vertex also vanish. In addition, the ultra-violet
divergent terms in the remaining diagram, the one with three 3-point vertices,
also vanish for the same reasons. This is analogous to what already happens
in the pure cubic theory \cite{BGI}.

The only surviving term is the infra-red divergent one in the last diagram 
above, which is proportional to
\begin{equation}
\kappa^{3}(c_{13}{\bar a}_{13})^{3}\int_{0}^{\infty} d\alpha_{1}d\alpha_{2}
d\alpha_{3} \frac{\alpha_{1}^{2}\alpha_{2}^{2}\alpha_{3}^{2}}
{(\alpha_{1}+\alpha_{2}+\alpha_{3})^{8}}\exp(-(\alpha_{1}+\alpha_{2}+
\alpha_{3})\epsilon) (\int d^{4}p \exp(ip^{2})).
\label{3.10}
\end{equation}
We stress that one should be careful in interpreting the integrals in
(\ref{3.10}), since it is not clear which regularizing prescription to use,
because of the peculiarities of the ${\rm (2,2)}$ signature. 

The Schwinger parameter integration in (\ref{3.10}) gives an infra-red 
divergence of the form 
$$
\int_{\varepsilon} ds \frac{1}{s^{3}}\sim\frac{1}{\varepsilon^{2}} 
$$
where $\varepsilon$ is some infra-red cut-off. This is indeed equivalent to 
the result of \cite{BGI}. However, one also has to note an infra-red divergence
associated with the singularity due to the ${\rm (2,2)}$ metric. In fact,
the gaussian momentum integral in (\ref{3.10}) is also divergent. In the 
present 
approach, two of the momentum components come from large $N$ color,
$(k_{q},k_{p})=(2\pi n_{q}/N, 2\pi n_{p}/N)$. This defines a natural 
infra-red regulator $\varepsilon =2\pi /N$. The transition from the sum over
$n_{q},n_{p}$ to the integral $\int dqdp$ then involves a factor of $N^{2}$
which is equivalent to $1/\varepsilon^{2}$. The total dependence on the 
infra-red regulator would then be 
$1/\varepsilon^{2}\cdot 1/\varepsilon^{2}=1/\varepsilon^{4}$ which would agree
with the $N=2$ string one. 

%%%%%%%%%%%%%%% SECTION 4 %%%%%%%%%%%%%%%%%

\section{Conclusions}

In this paper, we start the exploration of the quantum properties of
four-dimensional self-dual gravity viewed as a large $N$ two-dimensional 
WZW sigma 
model. We have shown how the expansion around a non-trivial classical 
background for the two-dimensional sigma model yields a four-dimensional 
field theory living in a space-time of signature ${\rm (2,2)}$.
In this approach, we generate an infinite sequence of higher point
vertices, which can be determined systematically from the two-dimensional
sigma model. At tree level, we have seen that our model is related to 
a particular dimensional reduction of four-dimensional large $N$ self-dual 
Yang-Mills theory. 
As such, it is closely related to a conjecture of Ooguri and Vafa 
(last paper of \cite{OV}).

We have checked that the amplitudes
of our model are consistent with the amplitudes of the closed $N=2$ string,
after we allow for a Lorentz $SO(2,2)$ transformation.
The 3-point amplitude at one-loop level in our model, is similar to
the one of the usual cubic actions for self-dual gravity, with
ultra-violet finiteness and an infra-red divergence that is weaker than the
one from the string. We believe that our approach may contain the 
solution to this puzzle, through a careful examination of the large $N$ 
limit or maybe through a more specific choice of Poisson algebra taking into 
account the global topology of the surface $\Sigma$. 
Although this naive argument should of course
be further substantiated, it provides an example of how a better agreement
between the field theory and the string could be achieved at the quantum 
level in this framework.
Most importantly, given the 
good quantum properties of the WZW model, we expect that the infinite series 
of terms in our lagrangian is likely to have important and desirable 
consequences at higher loop level.

It would be interesting to apply this large $N$ approach in the case where 
worldsheet instantons (from the Maxwell field in the $N=2$ 
worldsheet supergravity multiplet) are included in the string amplitudes 
\cite{OV}--\cite{L} and achieve a better understanding of the four-dimensional
Lorentzian properties of these field theories.

We note that a similar approach should also work in the case of the 
open string. Indeed, classical self-dual Yang-Mills theory can also be 
obtained from a large $N$ two-dimensional sigma model, where one can generate 
Chan-Paton factors or gauge groups in four dimensions if one extends the 
Poisson algebra $sdiff(\Sigma)$ by the Lie algebra of the chosen gauge group 
\cite{P}. Related considerations would allow the extension to the heterotic 
string as well \cite{OV,KM}.

%%%%%%%%%%%%%%%%%%%%%%%%%%%%%%%%%%%%%%%%%%%%%%%%%%%%%%%%%%%%%%%%%%%%%%%
%%%%%%%%%%%%%%%%%%% REFERENCES %%%%%%%%%%%%%%%%%%%%%%%%%%%%%%%%%%%%%%%%


\begin{thebibliography}{ZZZ}

\bibitem{dadda} M. Ademollo, L. Brink, A. D'Adda, R. D'Auria, E. Napolitano, 
S. Sciuto, E. Del Guido, P. di Vecchia, S. Ferrara, F. Gliozzi, R. Musto and
R. Pettorini, \PLB {\bf 62} (1976) 105.

\bibitem{dadda1} M. Ademollo, L. Brink, A. D'Adda, R. D'Auria, E. Napolitano, 
S. Sciuto, E. Del Guido, P. di Vecchia, S. Ferrara, F. Gliozzi, R. Musto,
R. Pettorini and J. Schwarz, \NPB {\bf 111} (1976) 77; \NPB {\bf 114} (1976) 
297.

\bibitem{MM} S. Mathur and S. Mukhi, \NPB {\bf 302} (1988) 130.

\bibitem{OV} H. Ooguri and C. Vafa, \MPL {\bf 5} (1990) 1389; \NPB {\bf 361} 
(1991) 469; \NPB {\bf 367} (1991) 83; H. Ooguri and C. Vafa, \NPB {\bf 451} 
(1995) 121, hep-th/9505183.

\bibitem{BGI} M. Bonini, E. Gava and R. Iengo \MPL {\bf 6} (1991) 795.

\bibitem{M} N. Marcus, \NPB {\bf 387} (1992) 263, hep-th/9207024;
``A Tour Through $N=2$ Strings'', Rome String Theory Workshop 1992, 
hep-th/9211059.

\bibitem{Si} W. Siegel, Phys. Rev. Lett. {\bf 69} (1992) 1493;
Phys. Rev. {\bf D46} (1992) 3235; N. Berkovits and W. Siegel, 
``Covariant Field Theory For Self-Dual Strings'', hep-th/9703154.

\bibitem{Ber} N. Berkovits and C. Vafa, \NPB {\bf 433} (1995) 123,
hep-th/9407190; N. Berkovits, \PLB {\bf 350} (1995) 28, hep-th/9412179;
\NPB {\bf 450} (1995) 90, hep-th/9503099.

\bibitem{L} O. Lechtenfeld, ``The Self-Dual Critical $N=2$ String'',
Talk at ``Int. Conference In Problems Of QFT'', Ukraine 1996,
hep-th/9607106; J. Bischoff and O. Lechtenfeld, \PLB {\bf 390} (1997) 153,
hep-th/9608196;
O. Lechtenfeld and W. Siegel, ``$N=2$ Worldsheet Instantons
Yield Cubic Self-Dual Yang-Mills'', hep-th/9704076.

\bibitem{J} A. Jevicki, ``Matrix Models, Open Strings And Quantization Of
Membranes'', Talk at the Argonne Duality Institute, June-96, hep-th/9607187.

\bibitem{KM} D. Kutasov and E. Martinec, \NPB {\bf 477} (1996) 652, 
hep-th/9602049; ``M-Branes And $N=2$ Strings'', hep-th/9612102;
D. Kutasov, E. Martinec and M. O'Loughlin, \NPB {\bf 477} (1996) 675;
E. Martinec, ``Matrix Theory And ${\rm (2,1)}$ Strings'', hep-th/9706194.

\bibitem{G} M. Green, \NPB {\bf 293} (1987) 593.

\bibitem{Ho} J. Hoppe, \PLB {\bf 215} (1988) 706;
B. de Wit, J. Hoppe and H. Nicolai, \NPB {\bf 305} (1988) 545;
D. Fairlie, P. Fletcher and C. Zachos, \PLB {\bf 218} (1989) 203;
J. Hoppe and P. Schaller, \PLB {\bf 237} (1990) 407.

\bibitem{FIT} E. G. Floratos, J. Iliopoulos, G. Tiktopoulos, \PLB {\bf 217} 
(1989) 285.

\bibitem{Pl} J. Pleba\'nski, Jour. Math. Phys. {\bf 16} (1975) 2394.

\bibitem{P} Q. Park, \IJM {\bf 7} (1992) 1415.

\bibitem{PA} A. Parkes, \PLB {\bf 286} (1992) 265.

\bibitem{H} V. Husain, Phys. Rev. Lett. {\bf 72} (1994) 800.

\bibitem{PP} C. Castro, Jour. Math. Phys. {\bf 34} (1993) 681; 
J. Pleba\'nski and  M. Przanowski, Phys. Lett. {\bf A212} (1996) 
22, hep-th/9605233; J. Pleba\'nski, M. Przanowski and H. Garc\'{\i}a-Compe\'an,
\MPL {\bf 11} (1996) 663.

\bibitem{FJ} B. Fridling and A. Jevicki, \PLB {\bf 134} (1984) 70;
E. Fradkin and A. Tseytlin, Ann. Phys. {\bf 162} (1985) 31;
C. Zachos and T. Curtright, ``The Paradigm Of Pseudo-Dual Chiral 
Models'', hep-th/9407044.

\bibitem{NaS} S. Donaldson, Proc. Lond. Math. Soc. {\bf 50} (1985) 1;
V. Nair and J. Schiff, \PLB {\bf 246} (1990) 423; \NPB {\bf 371}
(1992) 329; V. Nair, ``K\"ahler-Chern-Simons Theory'', Talk at Strings and 
Symmetries 91, Stonybrook, hep-th/9110042.

\bibitem{LMNS} A. Losev, G. Moore, N. Nekrasov, S. Shatashvili,
Nucl. Phys. proc. Suppl. {\bf 46} (1996) 130, hep-th/9509151; \NPB {\bf 484} 
(1997) 196, hep-th/9606082.

\bibitem{PW} A. Polyakov and P. Wigman, \PLB {\bf 131} (1983) 121.

\bibitem{LS} H. Leuwyler and M. Schifman, \IJM {\bf 7} (1992) 795.

\bibitem{Z} W. Zakrzewski, ``Low-Dimensional Sigma Models'', A. Hilger 1989.

\bibitem{PBR} A. Popov, M. Bordemann and H. R\"omer, \PLB {\bf 385} (1996) 63.





\end{thebibliography}
\end{document}